\documentclass[aps,PRL,twocolumn,superscriptaddress]{revtex4}
\usepackage{graphicx}% Include figure files

\usepackage{color}
\usepackage{latexsym}

\usepackage{xcolor}
\usepackage{hyperref}
\usepackage{amsmath,amssymb}
\usepackage{float}
\usepackage{bm}

\begin{document}
\draft
\def\ds{\displaystyle}
\title{ Non-diffracting states at exceptional points }
\author{Cem Yuce}
\email{cyuce@eskisehir.edu.tr}
\address{Department of Physics, Faculty of Science, Eskisehir Technical University, Turkey}

\author{Hamidreza Ramezani }
\address{Department of Physics and Astronomy, University of Texas Rio Grande Valley, Edinburg, TX 78539, USA}

\date{\today}
%\pacs{ }
\begin{abstract}
We propose to use exceptional points (EPs) to construct diffraction-free beam propagation and localized power oscillation in lattices. Specifically, here we propose two systems to utilize EPs for diffraction-free beam propagation, one in synthetic gauge lattices and the other one, in unidirectionally coupled resonators where each resonator individually is capable of creating orbital angular momentum beams (OAM). In the second system, we introduce the concept of robust and tunable OAM beam propagation in discrete lattices. We show that one can create robust OAM beams in an arbitrary number of sites of a photonic lattice. Furthermore, we report power oscillation at the EP of a non-Hermitian lattice. Our study widens the study and application of EPs in different photonic systems including the OAM beams and their associated dynamics in discrete lattices. 
\end{abstract}
\maketitle

{\it Introduction---} Diffraction management in photonics is a long-lasting problem that limits the application of beams and pulses especially in space communication, image forming, optical lithography, and electromagnetic tweezers \cite{Eisenberg_2000}. Diffraction management finds its root in many different system including disordered systems \cite{Segev_2013}, nonlinear systems \cite{ALFINITO_1996, Morandotti_2001}, and flat bands generated by symmetries \cite{Deng_2003,Yulin_2013,Ge_2015,Ramezani_2017,Biesenthal_2019}, bound state in continuum \cite{Plotnik_2011}, and optical vortex beams \cite{Shen_2019}. Specifically, the last one, namely optical vortex beams, is of extreme interest due to its proprieties to construct and explore fundamental theories associated with basic physical phenomena in the light-matter interaction, topological structures, and quantum nature of light which ultimately find its application in quantum information \cite{Molina_Terriza_2007}, optical tweezers \cite{Paterson_2001}, microscopy \cite{F_rhapter_2005} and imaging \cite{Tamburini_2006} to name a few.

Recently an effort has been made to reduce the size of the optical structures used to create orbital angular momentum (OAM) beams and make it possible to generate OAM on chip-scale \cite{Hayenga_2019, Zhang_2020, Ding_2020, Zheng_2020, Sroor_2020}. Despite all these achievements in the creation of OMA beams, there is no study on the dynamics of OAM beams in discrete lattices.

In this paper, we open a new direction in the study of dynamics of OAM beams in {\it discrete lattices} made of coupled individual elements where each element is capable of the creation of OAM beams separately. From now on we call such lattices {\it OAM-lattices}. Specifically, we are interested in having localized robust non-diffracting OAM beams in an arbitrary element or possibly at several elements of the OAM-lattice. To this end we introduce an abstract concept, namely continues family of eigenstate associated with one EP. This abstract concept allows us to construct more than one robust OAM-beam with different intensities in coupled optical elements. This can be generalized to create diffraction-free beam propagation.
\begin{figure}
	\includegraphics[width=1.\linewidth, angle=0]{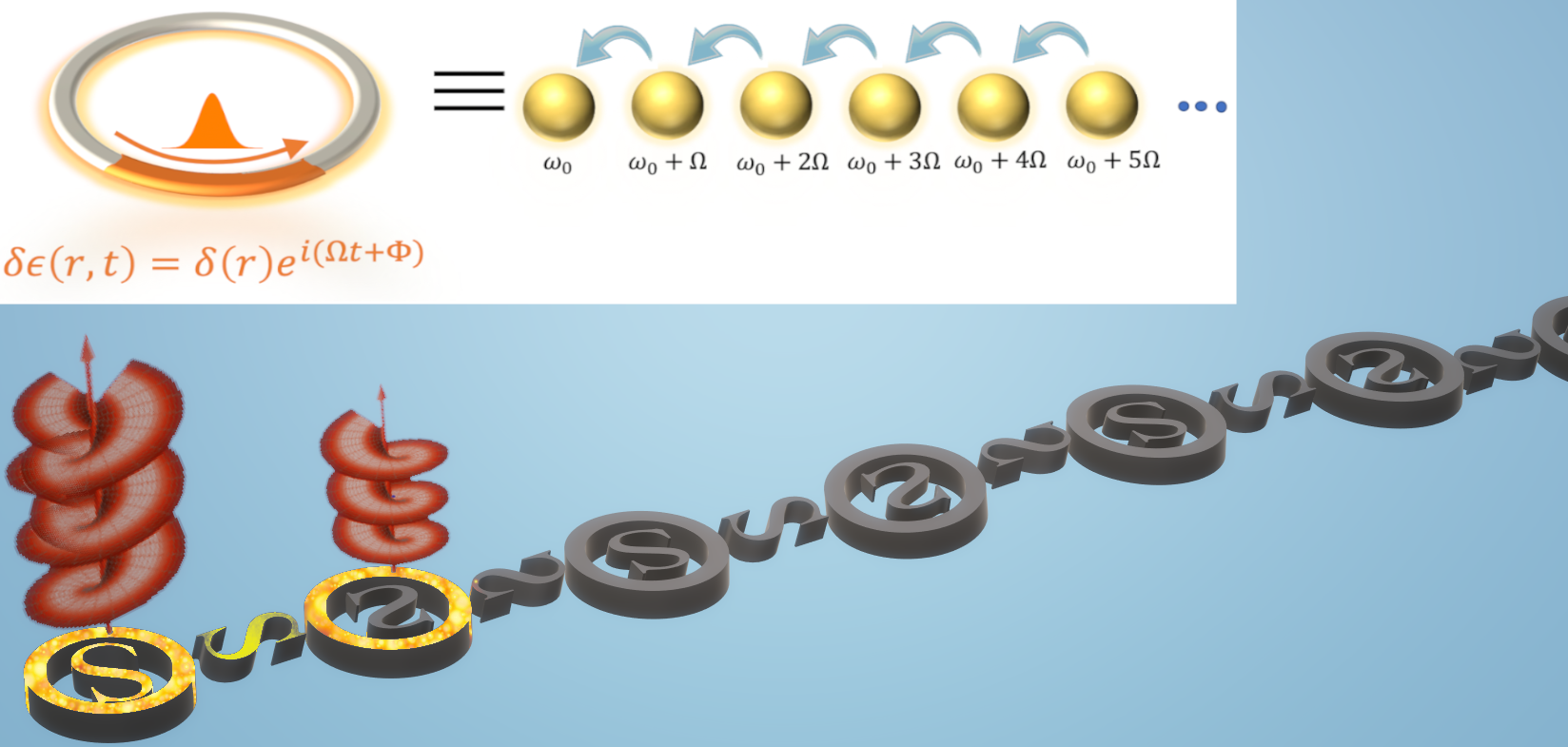}
	\caption{ (Main panel) Schematic of a semi-infinite OAM-lattice. The lattice made of microring resonators with enclosed S-shaped waveguides with tapered ends and S-shaped waveguides in between adjacent microrings. The tapered S-shaped waveguides cause a unidirectional coupling between the modal field amplitude of CCW and CW  in a microring or in between adjacent microrings. Notice that the face of the S-shaped waveguides is flipping from one microring to the other one. The flipping of S-shape in every other microring makes sure that the surviving CCW (CW) modal field amplitude in each microring couples to the CW (CCW) modal field of nearest neighbor microring. This OAM-lattice supports an arbitrary number of localized OAM beams and can be manipulated to have power oscillation. (upper left panel) A single-mode ring resonator that its permittivity is perturbed by a complex time-dependent modulation. This can resemble a semi-infinite synthetic lattice that its artificial sites are coupled to each other only in one direction. $\omega_0$ here denotes the resonance frequency of the static ring. One can use this approach to create a synthetic lattice for constructing non-diffracting beams based on the continuous family solution at EPs. }
	\label{fig1}
\end{figure}

EPs are topological singularities manifested in non-Hermitian systems \cite{Kato_1995}. By definition, at an $n$th order EP, $n$ eigenfunctions coalesce resulting to identical, well-defined, and unique eigenfunctions at a degenerate eigenvalue. EPs are ubiquitous in open systems and non-Hermitian optics and have interesting features such as unidirectional invisibility \cite{Lin_2011}, unidirectional lasing \cite{Ramezani_2014}, lasing and anti-lasing in a cavity \cite{Longhi_2010,Wong_2016}  and enhanced optical sensitivity \cite{Chen_2017}.

Here we reveal an abstract concept namely a family of EPs appearing in semi-infinite lattices that are defined by a continuous family of eigenstates. While such singular points occur at an infinite order EPs, and might not have direct physical realization, we use this class of EPs in two distinct models to construct non-diffracting beams in (i) artificial lattices and (ii) generate robust localized OAM beams in a finite OAM-lattice. 

{\it Model and discussion--}
Let us start with a semi-infinite photonic lattice made of coupled microring lasers each capable of vector vortex generation. Such microrings can be fabricated utilizing breaking rotation symmetry and asymmetric loss using an S-shape waveguide with adiabatically tapered ends embedded at the center of a microring \cite{Hayenga_2019}. A tapered S-shaped waveguide makes it possible to create a unidirectional coupling between two adjacent microrings by breaking the time-reversal symmetry and chirality. The orientation of the S-shaped waveguides is flipped in every other microring and in between the microrings as depicted in the upper panel of Fig(\ref{fig1}).

%\subsection{ Model and discussion}

Using the coupled-mode theory, we can write the modal field amplitudes in the clockwise and counterclockwise directions $(E_{CW},E_{CCW})^T_m=(\bar{\psi},\bar{\phi})^T_m$ ($T$ stands for transpose) at the $m(\in \text{odd})$th microring as \cite{Hayenga_2019}:
\begin{equation}\label{Eq1}
\begin{array}{c}

i\frac{d\bar{\psi}_m}{dt}=\sigma\bar{\psi}_m+w\bar{\phi}_{m+1}\\
i\frac{d\bar{\phi}m}{dt}=\sigma\bar{\phi}_m+v\bar{\psi}_{m}.

\end{array}
\end{equation}
The modal field amplitudes at the microrings with even $m$ are given by the same equation as the one in Eq.(\ref{Eq1}) except that we switch the $\bar\psi$ and $\bar{\phi}$ in Eq.(\ref{Eq1}). Here, for now we assume that $m$ starts from one and go to infinity, $t$ is time, $\sigma=-\omega_0+i(g-\gamma)$ is a constant depicted by the design of the structure, $\omega_0$ is the resonant frequency, $g$ is the linear gain, $\gamma$ represents linear dissipation due to structural/material losses or cavity decay, $v$ and $w$ signify the unidirectional coupling from the clockwise to the counter-clockwise modes in a microring and in between two adjacent microrings, respectively. 

Before proceeding further let us make a transformation  $(\bar{\psi},\bar{\phi})^T_m=(\psi,\phi)^T_m e^{-i\sigma t}$. Using this transformation the Hamiltonian associated with the $(\psi,\phi)$ of this semi-infinite OAM-lattice can be written as 
\begin{equation} \label{Eq2}
H_{EP}=\sum_{n=1}^{ 2N  }(c_1-(-1)^n c_2)|n\rangle\langle n+1|%+\sum_{n=1}^{\infty}\sigma|n\rangle\langle n|
\end{equation}
where $c_1 + c_2 = v$, $c_1 - c_2 = w$, $|n\rangle$ is a $\infty \times 1$ column vector that all of its elements are zero except the element at the $n$th row which is one, $N$ is the total number microrings, which is infinity for our system. This Hamiltonian is a Hamiltonian at EP as it is non-diagonalizable, featuring a $2N$-th order EP. The matrix form of $\ds{H_{EP}}$ is an upper triangular matrix with $\ds{(H_{EP})_{i,j} =0  }$ unless $\ds{  i\neq  j-1 }$ and hence it is a Jordan block matrix. At first glance, one would say that all eigenstates coalesce to a unique eigenvector
\begin{equation}\label{eql}
|\alpha\rangle=(1,{\bf 0})^T
\end{equation}
where $\bf 0$ is $1\times (2N-1)$, with the corresponding uniquely given eigenvalue $\alpha=0$. Here we show that this is not true for this semi-infinite lattice. Instead, the system admits a family of eigenstates with continuous eigenvalues.

At this point, we are interested in finding an analytical solution for the eigenvector associated with this EP. Let us expand the eigenvector $|\alpha\rangle$ of the Hamiltonian in Eq.(\ref{Eq2}) with its corresponding eigenvalue $\alpha$ in the $|n\rangle$-basis as
\begin{equation}\label{Eq3}
|\alpha\rangle=\sum_{n=1}^{\infty}   ~\xi_n  ~|n\rangle
\end{equation} 
where
\begin{equation}\label{Eq4}
H_{EP}~|\alpha\rangle=\alpha~ |\alpha\rangle.
\end{equation} 
Below we will show that the eigenvalue $\alpha$ of the Hamiltonian at the EP is a continues parameter. Substituting Eq.(\ref{Eq2}) and Eq.(\ref{Eq3}) in the Eq.(\ref{Eq4}) yields a recurrence equation for $\xi_n$. One can show that apart from a normalization constant, $\xi_n$ is given by 
\begin{equation}\label{Eq7}
\xi_n=\begin{cases}\begin{array}{ccc}
\alpha ^{n-1} (v w)^{\frac{1-n}{2}} & \text{if} & n\in \text{odd}\\
\alpha^{n-1} w(v w)^{-\frac{n}{2}} & \text{if} & n\in \text{even}\\
\end{array}
\end{cases}
\end{equation}
and thus the non-orthogonal EP eigenvector is given by
\begin{eqnarray}\label{Eq8}
\nonumber
|\alpha\rangle&=&\sum_{n\in \text{odd}} ^{\infty}\alpha ^{n-1} (v w)^{\frac{1-n}{2}}|n\rangle\\
&+&\sum_{n\in \text{even}}^{\infty}\alpha ^{n-1} w(v w)^{-\frac{n}{2}}|n\rangle
\end{eqnarray} 
forming continuous eigenvectors depending on the value of parameter $\alpha$. At this point the striking difference between the continuous family of solution at the EP in Eq.(\ref{Eq8}) and the conventional eigenvector at the EPs in Eq.(\ref{eql}) is clear. This continues family of eigenstates at EP arises for the semi-infinite system and has no analog when the system has finite number of sites, where all eigenstates coalesce. To study the eigenvalues $\alpha$ more specifically, it is instructive to see the expanded format of the eigenvector $|\alpha\rangle$ in Eq.(\ref{Eq8})
\begin{eqnarray}\label{Eq9}
\nonumber
|\alpha\rangle&=&|1\rangle+\frac{\alpha}{v}|2\rangle
+\frac{ \alpha^2}{v w}|3\rangle+\frac{\alpha^3}{v^2 w}|4\rangle\\&+&\frac{\alpha^4}{v^2 w^2}|5\rangle+\frac{\alpha^5}{v^3 w^2}|6\rangle+....
\end{eqnarray}
The solution Eq.(\ref{Eq3}) is normalizable if $\ds{\xi_{n\rightarrow\infty}  \rightarrow 0 }$. One can see that this condition is satisfied if $\ds{|\frac{\alpha}{v}|<1 }$ and $\ds{|\frac{\alpha^2}{vw}|<1 }$. Therefore $\ds{\alpha}$ can be a real or a complex number which seems to be in contrast to the generally believed one where an EP determines a phase transition from a real spectrum to a complex spectrum. Specifically when $v=w$ the total intensity has a closed form $\langle\alpha|\alpha\rangle=(1-|\frac{\alpha}{v}|^2)^{-1}$ which obviously indicate why $|\frac{\alpha}{v}|<1$. By varying $\alpha$ at fixed $v$ and $w$, one can obtain a continuous family of square integrable and stable OAM beams. Notice that $\ds{\xi_{n}  }$ rapidly goes to zero with $n$ for small values of $|\alpha|$. This implies that our solution, which seems valid only for the semi-infinite lattice, can still be used to construct almost stationary states for a truncated lattice. In other words, $|\alpha\rangle$ becomes 
an effectively stationary state for the truncated lattice at the EP in the timescale of an experiment for sufficiently small values of $\alpha$. As we discuss later, depending on the value of couplings we can have different scenarios where OAM beams with different amplitude leave in several microrings. %One can chose parameters such that somewhere inside the lattice one of the CCW or CW modes are significantly larger than the other one and has an almost pure OAM beam inside the lattice while other microrings before this microring have a combination of CW and CCW modes with different intensities. 
%We would like to mention that adding disorder in couplings $v$ and $w$ does not affect this solution meaning that the OAM beams are robust against disorder in the couplings $v,w$. 
Furthermore, $\alpha$ and coupling values $v$ and $w$ dictate how many OAM beams remain localized at the left side of the lattice and at the limit of $\alpha\to 0$ only one pure OAM beam will be localized at the first resonator. Notice that $\alpha=0$ reduces to the solution given in Eq.(\ref{eql}). Physically, choosing different values for $\alpha$ means that one can construct arbitrary numbers of OAM beams in the lattice just by adjusting their relative intensities.% and interestingly enough always remain at the EP irrespective of the value of $\alpha$.

\begin{figure}[t]
	\includegraphics[width=\columnwidth]{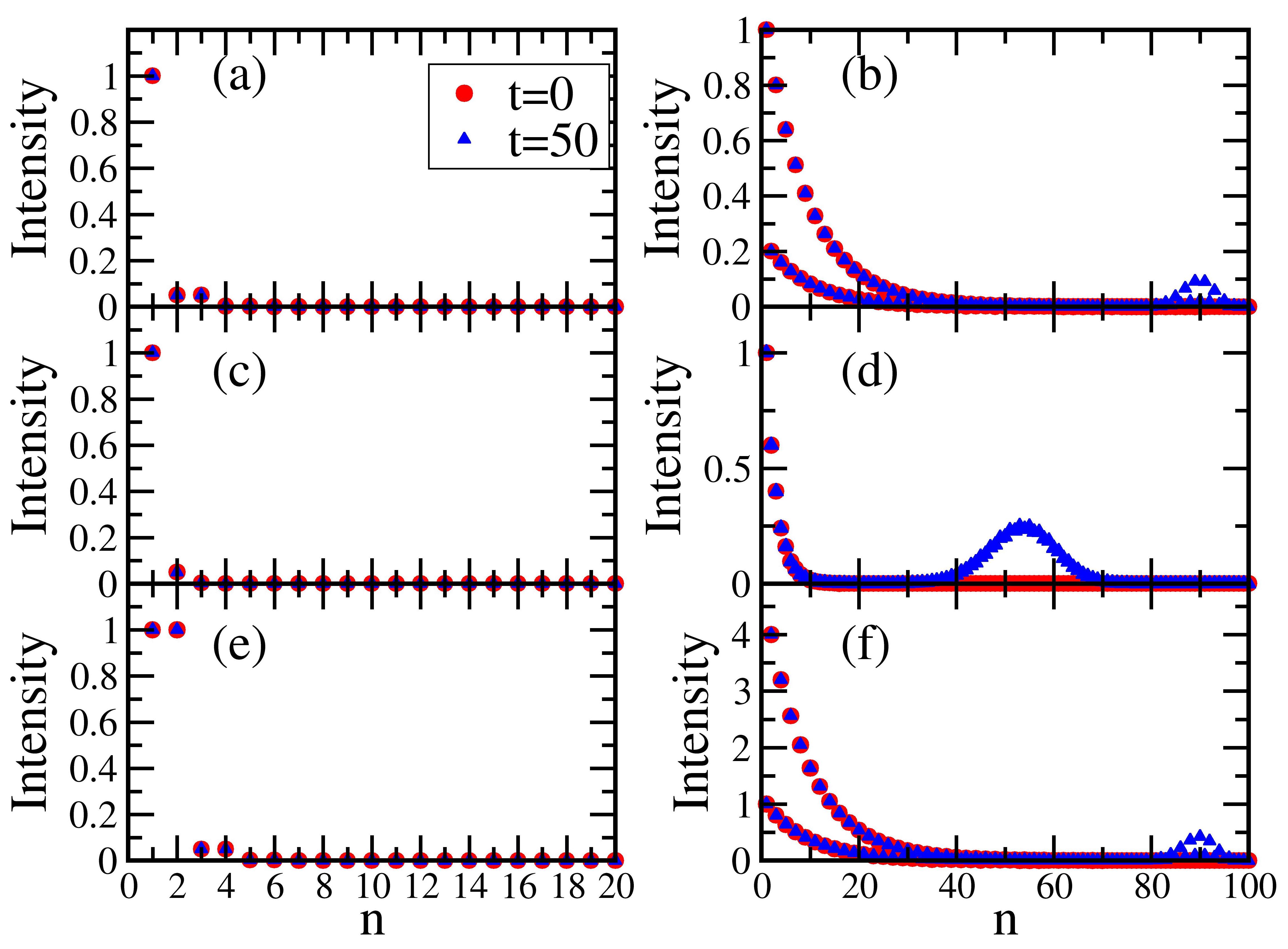}
	\caption{ Diffraction behavior of the initial excitation of a finite size OAM-lattice associated with an upper panel of Fig. (\ref{fig1}) which is composed of 50 microrings (in the left column we just show 10 rings, or 20 first modal amplitudes $n$ at the left side of the lattice as the rest are almost zero) for different values of coupling and $\epsilon$ at time $t=0$ (red circle) and $t=50$ (blue triangle) in the unit of coupling. We assumed $\sigma=0$ for simplicity. In (a,b) Coupling $w=0.05$ is much smaller than $v=1$, $\alpha=0.05$ in (a) and $\alpha=0.2$ in (b). In this case $t$ is normalized w.r.t. $v$. As expected in (a) a dominant CCW OAM beam leaves in the first resonator with some small contribution from a CW mode. The second microring has mainly CW mode with a smaller amplitude. In (b) the larger $\alpha$ is equivalent to the excitation of more ring resonators. In this case, the effect of finite size lattice becomes apparent mostly at the left of the lattice where a pick appears and starts to grow exponentially. In (c,d) the coupling $v=1$ and $w=0.9$ and t is again normalized to $v$. Furthermore, we picked $\alpha=0.05$ for (c) and $\alpha=0.6$ for (d). Again the same behavior as (a,b) is observed with different intensities at the left side of the lattice in each case. In (e,f) we made $v=0.05$ and $w=1$ and normalized $t$ w.r.t. $w$. Interestingly enough in (e) there is no pure CCW or CW mode in any ring and both of them have similar intensity in a ring. It is clear that by choosing the couplings and different values of $\alpha$ we can make different localized beam propagation with almost no diffraction.}
	\label{fig2}
\end{figure}
\begin{figure}[t]
	\includegraphics[width=4cm]{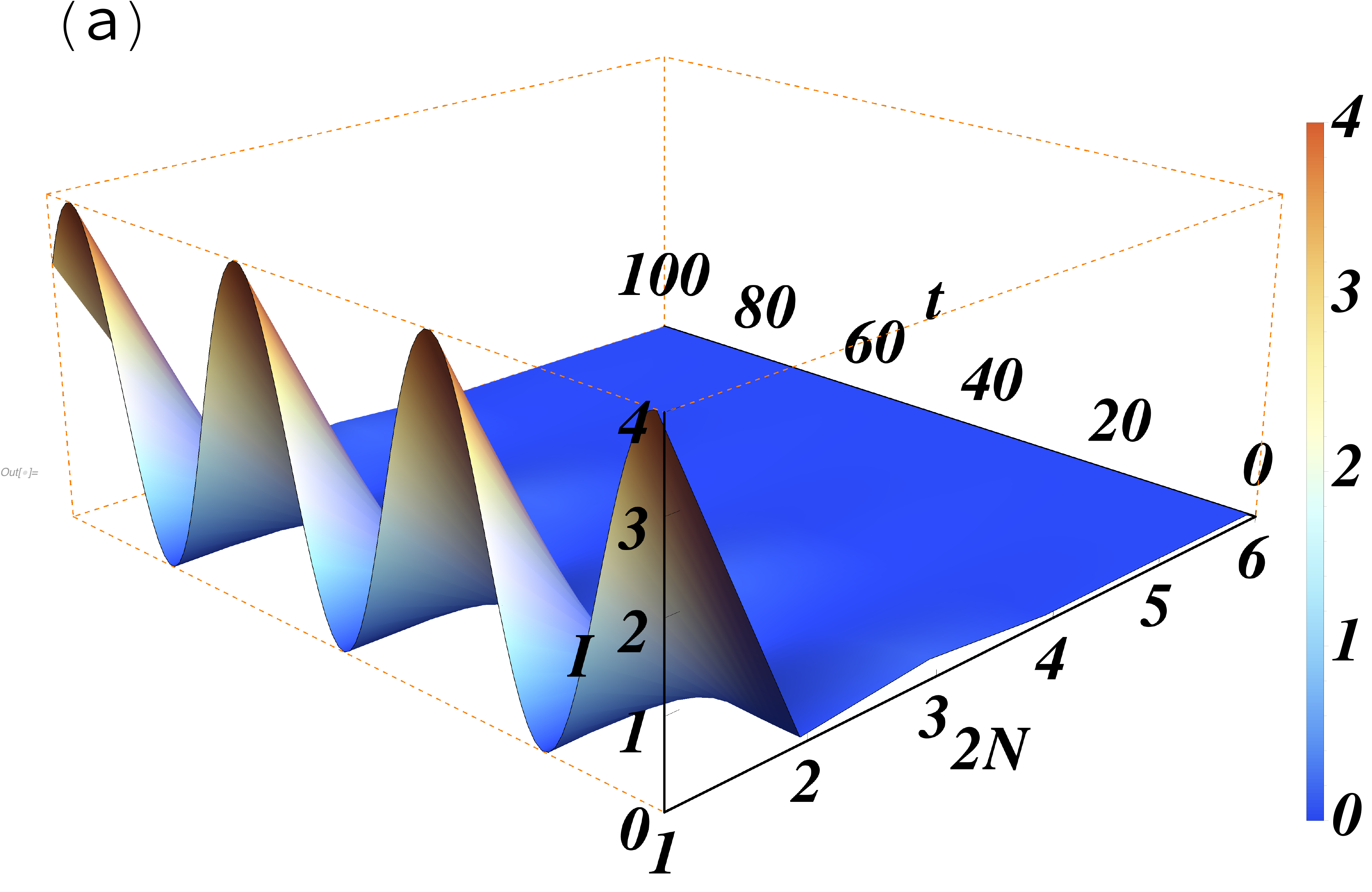}
	\includegraphics[width=4cm]{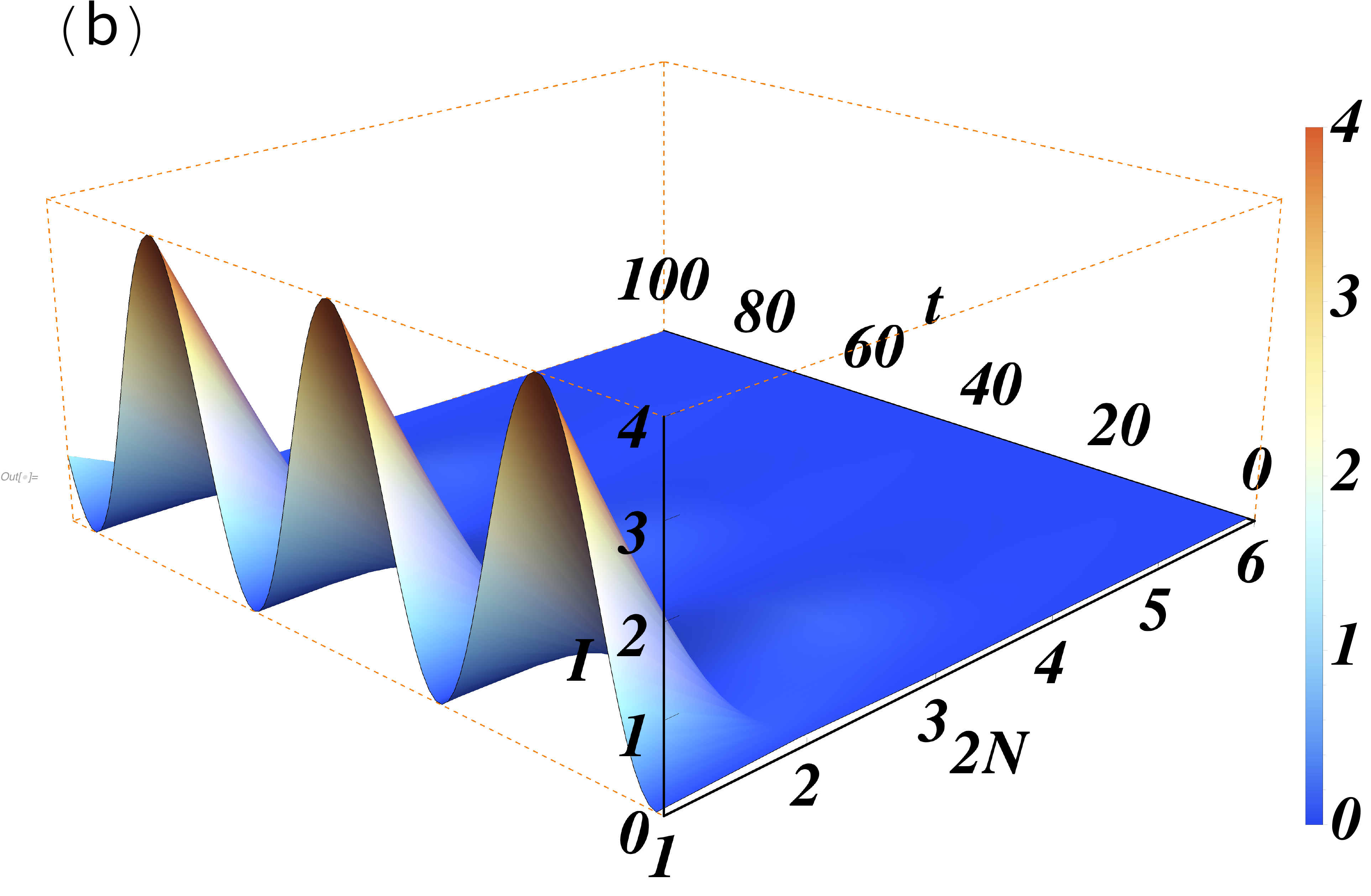}
	\includegraphics[width=4cm]{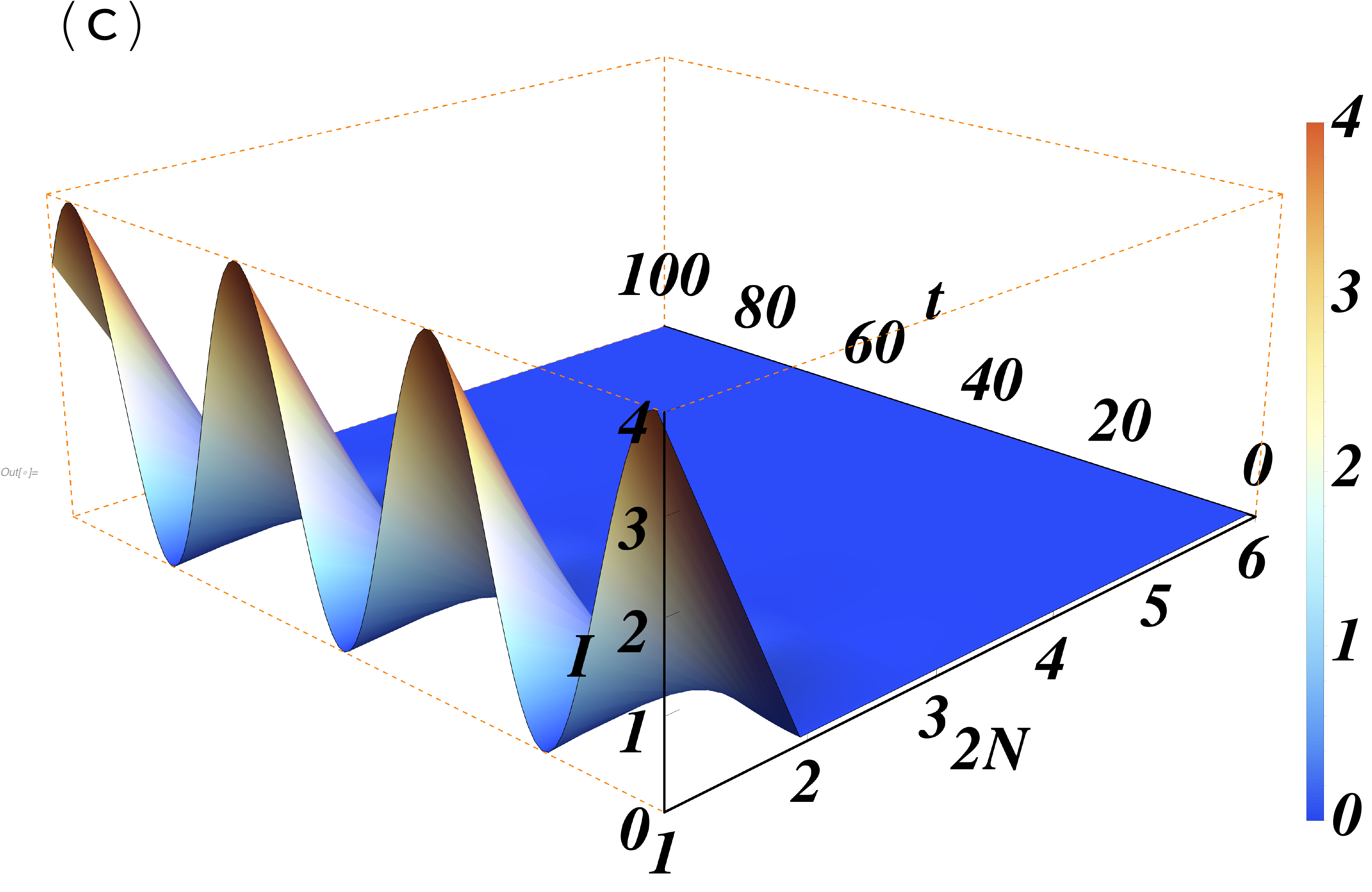}
	\includegraphics[width=4cm]{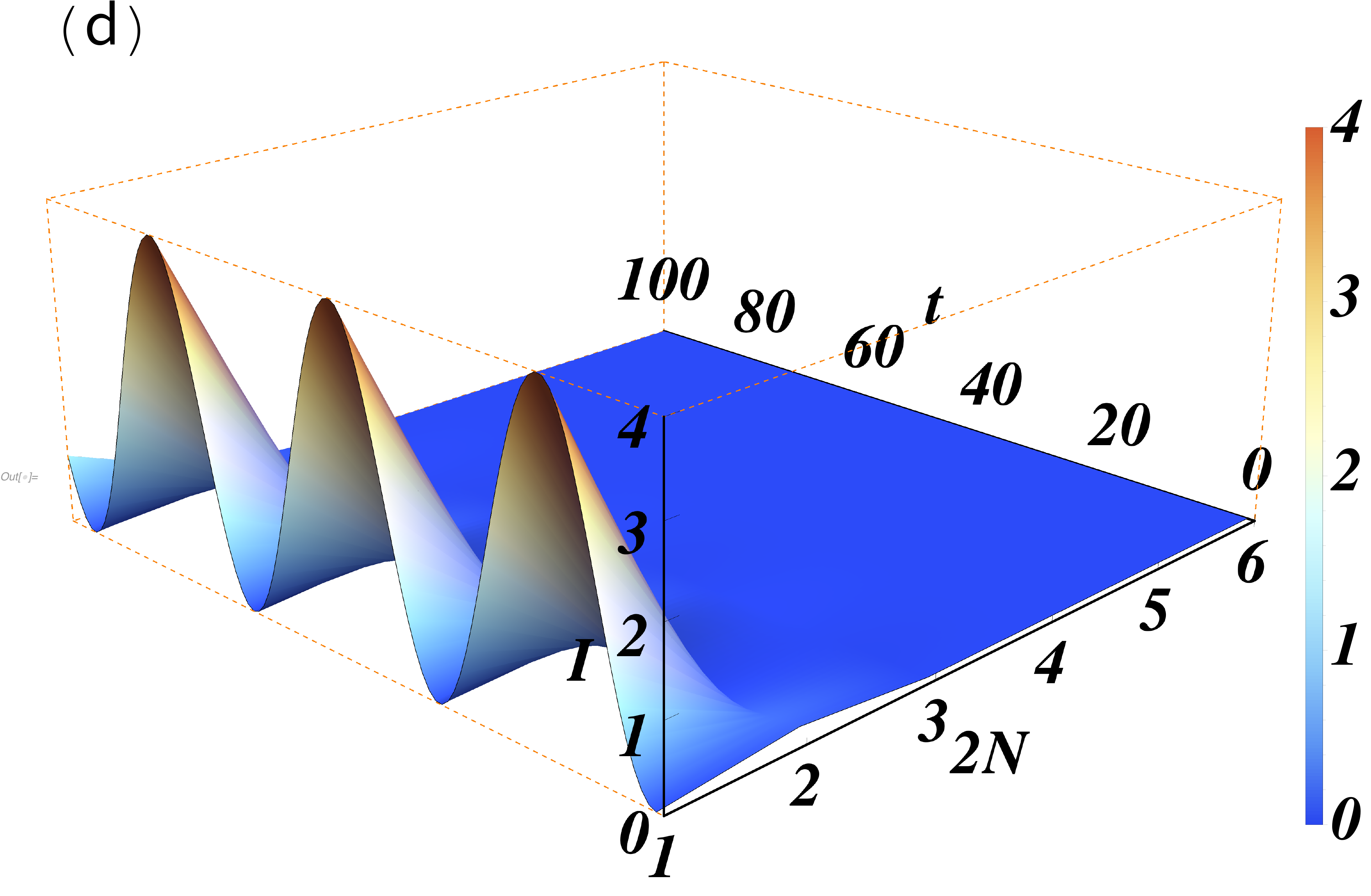}
	
	\caption{ Dynamics of the OAM beams in a lattice with 50 ring resonators with initial conditions given by the superposition of $\alpha=\pm0.1$ in Eq.(\ref{Eq9}). The left column is associated with the case where we use the plus sign for the superposition while the left column is for minus superposition. In (a,b) $v=1$, $w=0.05$, and (c,d) $v=0.5$, $w=1$. $t$ is the time in the unit of the coupling which has value one, $N$ is the number of resonators, only modes of three rings which have significant intensity are shown here, and $I$ is the intensity of the corresponding mode. We observe a power oscillation between the CCW modes and CW modes. One can control the relative intensity just by adjusting the couplings.}
	\label{soam}
\end{figure}

The solution in Eq.(\ref{Eq7}) is only valid for the semi-infinite lattice and thus one might become suspicious about its practical application which needs large lattice size. To address this issue and before discussing further the effect of lattice size  and also couplings on the OAM beams here we propose an approach to build a very large lattice in the synthetic dimension. Creating lattices in the synthetic dimension with bidirectional coupling has been studied before where one uses a modulation defined via a pure real function to create an artificial dimension, such as frequency for instance, and make a synthetic lattice \cite{Ozawa_2016,Luo_2017}. In contrast to what has been proposed so far here we are interested in a large lattice with artificial sites that its artificial sites are coupled in a unidirectional manner as needed for the observation of the continues family of solutions at the EP \cite{Li_2018}. For this purpose consider, for example, a single mode ring resonator as shown in the upper corner of the Fig.(\ref{fig1}). Let us assume that the permittivity of the ring is perturbed with a complex function 
\begin{equation}
\delta \epsilon(r,t)=\delta(r) e^{i(\Omega t+\Phi)}
\end{equation} 
where $\delta(r)$ is the modulation profile, $\Omega$ is the modulation frequency, and $\Phi$ is the modulation phase. Notice that unlike the previous proposals on synthetic lattices we are using a complex function modulation rather than a real function which means that we are modulating not only the real part of the permittivity but also its imaginary part which physically is equivalent to time dependent gain or loss mechanism. Such modulation forms a list of equally spaced modes started at the resonance frequency of the static ring resonator where the dynamics of these modes are given by the following equation:
\begin{equation}\label{eq11}
i\frac{da_m}{dt}=\kappa e^{-i\Phi} a_{m+1}
\end{equation}
where $\kappa$ is the coupling strength between the modes. For the detailed derivation of Eq.(\ref{eq11}) see Ref.(\cite{Yuan_2016, Ozawa_2016}). One can see the immediate connection between the Eq.(\ref{eq11}) and Eq.(\ref{Eq1}).

It is easy to show that by truncating the lattice the EP occurs at $\alpha=0$ with a corresponding eigenvector given in Eq.(\ref{eql}). Our numerical simulations shows that for a lattice which is long enough still one can excite the lattice with an initial condition given by Eq.(\ref{Eq8}) and have non-diffracting OAM beams for a large period of time. As the lattice size becomes smaller, one is forced to use a smaller value for $\alpha$ to observe diffraction free beams. %Furthermore, if $\alpha^q\gg\alpha^{q+1}$ where $q$ is an even integer then in the ring number $q$ one has almost a pure OAM beam that is not diffracting while other previous rings have a combination of OAM beams with positive and negative angular momentum that have uneven intensities. 

Next let us discuss the effect of couplings and value of $\alpha$ on the dynamics in a finite size lattice. As depicted in Fig.(\ref{fig2}) by choosing different values for couplings and $\alpha$ we can create different OAM beams that are localized in the left side of the lattice. One can have a combination of OAM beams with different amplitudes in a ring or a single OAM beam in a microring or a combination of them where first there is mixture of them on the left and eventually there is a pure OAM beam in ring. The whole process is almost with no diffraction for a long  time $t$. This picture disturbs when the pick which is much down in the lattice reaches to most left microring.

The linear nature of the Eq.(\ref{Eq1}) and solution (\ref{Eq9}) allows us to further manipulate the dynamics of the generated OAM beams. For example, one can excite the lattice with the superposition of $|\alpha_1\rangle$ and $|\alpha_2\rangle$ such that initial excitation is given by $|\alpha_1\rangle \pm |\alpha_2\rangle$ and with $\alpha_1=-\alpha_2=\alpha\in \Re$. It is straightforward to see that the aforementioned superposition with the plus sign results in $ |\alpha\rangle\propto|1\rangle
+\frac{ \alpha^2}{v w}|3\rangle+\frac{\alpha^4}{v^2 w^2}|5\rangle+....$ which equivalent of exciting the CCW and CW modes in every other microring. On the other hand,  the subtraction results in $|\alpha\rangle\propto\frac{\alpha}{v}|2\rangle
+\frac{\alpha^3}{v^2 w}|4\rangle+\frac{\alpha^5}{v^3 w^2}|6\rangle+....$ which is equivalent to exciting CW and CCW modes in every other microring. As a result of such excitation, we can have only one type of OAM beams in each ring. In Fig.(\ref{soam}) we depicted the dynamics for different values of couplings and the aforementioned superposition of $\alpha$ and $-\alpha$ without any normalization of the initial excitation. We observe a surprising effect namely, a localization with power oscillation at an EP. Power oscillation has been observed in early studies of PT-symmetric systems where the bi-orthogonality causes the power oscillation\cite{Makris_2008, Musslimani_2008}. However, to the best of our knowledge, there is no report on the localized beam that has power oscillation at an EP. Due to the power oscillation, a CCW mode transfers to a CW mode and vise versa.  Specifically, for the plus superposition as shown in the left column of Fig.(\ref{soam}) initially the CCW modes with odd mode numbers depicted in the x-axis of Fig.(\ref{soam}) are excited. After some propagation time, $t$, the field intensity in the CCW modes decays and some power transfers to the CW mode. Again after a few coupling time propagation, the field intensity comes back to its initial values. Similar dynamics occur for the minus superposition where originally the even modes (CW modes) are excited. This power oscillation can be understood when we realize that the state $|\alpha(t)\rangle=\exp(i\alpha t)|\alpha\rangle\pm\exp(-i\alpha t)|-\alpha\rangle$.
Clearly one can control the intensity profile by detuning couplings $v$ and $w$. Notice that one expects to have a flow of intensity from higher mode numbers toward the mode one. However, as we see here the intensity oscillates between the modes due to the superposition.

As a final comment, while the transpose of the Hamiltonian in Eq.(\ref{Eq2}), which is equivalent to a lattice that its coupling is in the opposite direction as the one shown in the upper panel of Fig.(\ref{fig1}), has the same spectrum as the one in here, however, the system does not have a localized solution at the EP.
{\it Conclusion--}
In Conclusion, we have introduced continuous family of eigenstates at EPs in semi-infinite lattices. %This is not in agreement with the general belief that all eigenstates coalesce at EPs. 
This class of EP is a function of a free arbitrary parameter. By varying the free parameter, we get another eigenstate at the EP. Physically the changes in the free parameter are equivalent to the number of the excited state in the lattice. While this class of solution at EP only occurs in semi-infinite lattices we proposed two different systems to make use of this
family of solution at EPs for diffraction-free propagation, (i) in the frequency domain known as synthetic gauge lattices and (ii) in a spatial domain where we used microring resonator capable of generating OAM beams individually. In practice coupled microrings that are capable of making OAM beams disturb the generated OAM beams and thus so far there is no study on a lattice that each element can create OAM beam. Here we used our proposed EPs to study OAM beams in lattices and manipulate OAM beams in such lattices. Furthermore we reported a localized power oscillation at the EP. This localized power oscillation, apart from its localized feature, is a distinct property of our proposed EPs and is different from power oscillation in PT-symmetric systems where only in the exact phase one can see such power oscillation due to the bi-orthogonality.

%Despite the aforementioned extensive studies on Glauber-FocK states, there is no 
\begin{acknowledgments}
H.R acknowledge the support by the Army Research Office Grant No. W911NF-20-1-0276 and NSF Grant No. PHY-2012172. The views and conclusions contained in this document are those of the authors and should not be interpreted as representing the official policies, either expressed or implied, of the Army Research Office or the U.S. Government. The U.S. Government is authorized to reproduce and distribute reprints for Government purposes notwithstanding any copyright notation herein. 

\end{acknowledgments}

\end{document}